\documentclass[12pt]{article}

\newcommand{\be}{\begin{equation}}
\newcommand{\ee}{\end{equation}}
\newcommand{\bi}[1]{\vspace{-3mm} \bibitem{#1}}

\usepackage{epsfig,amsmath} 
\usepackage{graphics}

\begin{document}
\begin{center}
Modern Physics Letters B. Vol.19. No.15 (2005) pp.721-728.
\end{center}

\begin{center}
{\Large \bf Wave Equation for Fractal Solid String}
\vskip 5 mm

{\large \bf Vasily E. Tarasov } \\

\vskip 3mm
{\it Skobeltsyn Institute of Nuclear Physics, \\
Moscow State University, Moscow 119992, Russia}

{E-mail: tarasov@theory.sinp.msu.ru}
\end{center}
\vskip 11 mm

\begin{abstract}
We use the fractional integrals to describe fractal solid.
We suggest to consider the fractal solid as special 
(fractional) continuous medium.  
We replace the fractal solid with fractal mass dimension 
by some continuous model that is
described by fractional integrals. 
The fractional integrals are considered as approximation of 
the integrals on fractals. 
We derive fractional generalization of the action functional and
the Euler-Lagrange equation for the fractal solid string.
The solution of wave equation for fractal solid string
is considered.  
\end{abstract}



\section{Introduction}

The real fractal structures of matter are characterized 
by an extremely complex and irregular geometry \cite{Mand}. 
Although the fractal dimensionality does
not reflect completely the geometric 
properties of the fractal solid, it nevertheless permits
a number of important conclusions about the behavior 
of fractal structures \cite{Mand,Zaslavsky1,Zaslavsky2,Zaslavsky3}. 
For example, if it is assumed that matter with a constant density 
is distributed over the fractal, then the mass of the fractal 
enclosed in a volume of characteristic dimension
$R$ satisfies the scaling law $M (R) \sim R^{D}$, whereas
for a regular n-dimensional Euclidean object $M (R) \sim R^n$ \cite{Mand}.
Let us assume that a solid can be treated on a scale $R$ as 
a stochastic fractal of dimensionality $D <3$ embedded in a Euclidean
space of dimensionality $n = 3$. Naturally, in real objects 
the fractal structure cannot be observed on all scales.  
For example, Katz and Thompson \cite{KT} presented experimental 
evidence indicating that the pore spaces of a set of
sandstone samples are fractals. 

In the general case, the fractal solid cannot 
be considered as a continuous medium.
There are domains which are not filled by particles.
We suggest \cite{PLA05,AP05-2,PLA05-2} to consider the fractal solid  
as special (fractional) continuous solid.
We use the procedure of replacement of the fractal solid 
with fractal mass dimension by some continuous solid that 
is described by fractional integrals.
This procedure is a fractional generalization of 
Christensen approach \cite{Chr}.
Suggested procedure leads to the fractional integration and 
differentiation to describe fractal solid.
The fractional integrals allow us to take into account the fractality
of the solid \cite{PLA05,AP05-2}.
In order to describe the fractal solid by continuous model
we must use the fractional integrals \cite{SKM,FA}. 
The order of fractional integral is equal 
to the fractal mass dimension of the solid.
More consistent approach to describe the fractal solid
is connected with the mathematical definition of the integrals on fractals. 
It was proved \cite{RLWQ} that integrals on net of fractals 
can be approximated by fractional integrals. 
Therefore, we can consider the fractional integrals as approximations  
of integrals on fractals \cite{Nig4}.
In Ref. \cite{chaos,PRE05,JPCS}, we proved that fractional integrals 
can be considered as integrals over the space with fractional 
dimension up to numerical factor. To prove this statement, 
we use the well-known formulas of dimensional regularizations \cite{Col}.  

In this paper, we use the fractional integrals in order
to describe dynamical processes in the fractal solid. 
In Sec. 2,  we consider the fractional continuous 
model for the fractal solid.
In Sec. 3, we derive the fractional generalization of 
the stationary action principle and the Euler-Lagrange equation. 
In Sec. 4, we consider the solution of 
wave equation for the fractal solid string.
Finally, a short conclusion is given in Sec. 5.

\section{Mass of Fractal Solid}

Let us consider a fractal solid.
For example, we can assume that mass 
with a constant density is distributed over the fractal. 
In this case, the mass $M(W)$ on the fractal enclosed 
in volume with the characteristic size $R$ satisfies the 
scaling law $M(R) \sim R^{D}$, whereas for a regular n-dimensional 
Euclidean object we have $M(R)\sim R^n$. 

The total mass in a region $W$ is given by the integral
\[ M(W)=\int_W \rho({\bf r},t) dV_3 . \]
The fractional generalization of this equation can be written
in the following form
\[ M(W)=\int_W \rho({\bf r},t) dV_D , \]
where $D$ is a mass fractal dimension of solid,
and $dV_D$ is an element of $D$-dimensional volume such that
\be \label{5a} dV_D=c_3(D,{\bf r})dV_3. \ee

For the Riesz definition of the fractional integral \cite{SKM}, the 
function $c_3(D,{\bf r})$ is defined by the relation
\be \label{5R} c_3(D,{\bf r})=
\frac{2^{3-D}\Gamma(3/2)}{\Gamma(D/2)} |{\bf r}|^{D-3} . \ee
The initial points of the fractional integral are set to zero \cite{SKM}.
The numerical factor in Eq. (\ref{5R}) has this form in order to
derive usual integral in the limit $D\rightarrow (3-0)$.
Note that the usual numerical factor
$\gamma^{-1}_3(D)=\Gamma(1/2)/(2^D \pi^{3/2} \Gamma(D/2))$,
which is used in Ref. \cite{SKM}
leads to $\gamma^{-1}_3(3-0)= \Gamma(1/2)/(2^3 \pi^{3/2}
\Gamma(3/2))=1/(4\pi^{3/2})$ in the limit $D\rightarrow (3-0)$.

For the Riemann-Liouville fractional integral \cite{SKM}, 
the function $c_3(D,{\bf r})$ is defined by
\be \label{5RL} c_3(D,{\bf r})=
\frac{|x y z |^{D/3-1}}{\Gamma^3(D/3)}  . \ee
Here we use Cartesian's coordinates $x$, $y$, and $z$. 
In order to have the usual dimensions of the physical values,
we can use vector ${\bf r}$, and coordinates 
$x$, $y$, $z$ as dimensionless values.

Note that the interpretation of the fractional integration
is connected with fractional dimension \cite{chaos,PRE05}.
This interpretation follows from the well-known formulas 
for dimensional regularizations \cite{Col}.
The fractional integral can be considered as an
integral in the fractional dimension space
up to the numerical factor $\Gamma(D/2) /( 2 \pi^{D/2} \Gamma(D))$.

If we consider the ball region $W=\{{\bf r}: \ |{\bf r}|\le R \}$, 
and spherically symmetric solid 
($\rho({\bf r},t)=\rho(r)$), then we have
\[ M(R)=4\pi \frac{2^{3-D}\Gamma(3/2)}{\Gamma(D/2)}
\int^R_0 \rho(r) r^{D-1} dr . \]
For the homogeneous ($\rho(r)=\rho_0$) fractal distribution, we get
\[ M(R)=4\pi \rho_0 \frac{2^{3-D}\Gamma(3/2)}{D\Gamma(D/2)}
R^D \sim R^D . \]
Fractal solid is called a homogeneous 
fractal solid if the power law  $M(R)\sim R^D $ does not 
depends on the translation  of the region. 
The homogeneity property of fractal solid
can be formulated in the following form:
For all regions $W$ and $W^{\prime}$ of the homogeneous fractal 
solid such that the volumes are equal $V(W)=V(W^{\prime})$, 
we have that the masses of these regions 
are equal $M(W)=M(W^{\prime})$. 
Note that the wide class of fractal media satisfies 
the homogeneous property.
In many cases, we can consider the porous media \cite{Por1,Por2}, 
polymers \cite{P}, colloid aggregates \cite{CA}, and 
aerogels \cite{aero} as homogeneous fractal media.
In Refs. \cite{PLA05,AP05-2}, the continuous medium model of the fractal 
media was suggested. Note that the fractality and 
homogeneity properties can be considered in the following forms: 

\noindent
(1) Homogeneity:
The local density of homogeneous fractal solid
is a translation invariant value that has the form
$\rho({\bf r})=\rho_0=const$.

\noindent
(2) Fractality:
The total mass of the ball region $W$ obeys a power law relation
$M(W) \sim R^D$,  where $D<3$, and $R$ is the radius of the ball.

\section{Fractional Action Functional for Fractal Solid String}

Let us consider the string that is described by the field $u(x,t)$, 
where $0<x<l$. 
Suppose that the density of this string is constant $\rho(x,t)=\rho_0=const$. 
Further assume that the string is stretched with the constant force $F_0$.
The kinetic energy of this string is defined by the equation
\be \label{T}
T=\frac{1}{2} \rho_0 \int^l_0 
\Bigl( \frac{\partial u}{\partial t} \Bigr)^2 d x .
\ee  
The fractional generalization of this equation has the form 
\be \label{TD}
T=\frac{1}{2} \rho_0 \int^l_0 
\Bigl( \frac{\partial u}{\partial t} \Bigr)^2 dl_D ,
\ee  
where we use the following notations
\be \label{lD} dl_D=c_1(D,x)dx , \quad 
c_1(D,x)=\frac{|x|^{D-1}}{\Gamma(D)} . \ee
Equation (\ref{TD}) defines the kinetic energy of the homogeneous fractal
solid string. 
If the oscillation has the small amplitude, then the potential energy
can be represented by the integral
\be \label{U} 
U=\frac{1}{2}F_0 \int^l_0 \Bigl( \frac{\partial u}{\partial x} \Bigr)^2 d x .
\ee
The fractional generalization of this equation has the form 
\be \label{UD} 
U=\frac{1}{2}F_0 \int^l_0 \Bigl( \frac{\partial u}{\partial x} \Bigr)^2 dl_D .
\ee

The Lagrange function is defined by the equation $L=T-U$.
The action for the time interval $0\le t \le \tau$ has the following form
\be
S[u]=\frac{1}{2} \int^{\tau}_0 dt
\int^l_0 \Bigl( \rho_0 \Bigl( \frac{\partial u}{\partial t} \Bigr)^2 -
F_0  \Bigl( \frac{\partial u}{\partial x} \Bigr)^2  \Bigr) dl_D .
\ee
The stationary action principle leads to the equation
\[ \Bigl( \frac{d}{d\varepsilon} 
S[u+\varepsilon \varphi]\Bigr)_{\varepsilon=0}=0 . \]
For the fractal solid string, we have
\[ \Bigl( \frac{d}{d\varepsilon} S[u+\varepsilon \varphi]\Bigr)_{\varepsilon=0}=
\frac{1}{2} \int^{\tau}_0 dt
\int^l_0 \Bigl( \rho_0 \frac{\partial u}{\partial t} 
\frac{\partial \varphi}{\partial t} -
F_0  \frac{\partial u}{\partial x} 
\frac{\partial \varphi}{\partial x} \Bigr) dl_D . \]
Using Eq. (\ref{lD}), the integration by part, and the conditions
\[ \varphi(x,0)=\varphi(x,\tau)=0, \quad  \varphi(0,t)=\varphi(l,t)=0 , \]
we get the equation
\be \label{sap2}
\Bigl( \frac{d}{d\varepsilon} S[u+\varepsilon \varphi]\Bigr)_{\varepsilon=0}
= -\frac{1}{2} \int^{\tau}_0 dt
\int^l_0 \Bigl( \rho_0 c_1(D,x) \frac{\partial^2 u}{\partial t^2} - F_0 
\frac{\partial}{\partial x} \Bigl( c_1(D,x) 
\frac{\partial u}{\partial x} \Bigr) \Bigr)  \varphi dx . \ee
The right-hand side of Eq. (\ref{sap2}) must be equal to zero for
all functions $\varphi(x,t)$. Therefore the following equation
must be satisfied
\be  
\rho_0 c_1(D,x) \frac{\partial^2 u}{\partial t^2} - F_0 
\frac{\partial}{\partial x} \Bigl( c_1(D,x) 
\frac{\partial u}{\partial x} \Bigr) =0 . \ee 
This Euler-Lagrange equation can be rewritten in an equivalent form
\be c_1(D,x) \frac{\partial^2 u}{\partial t^2} - 
v^2 \frac{\partial}{\partial x} \Bigl( c_1(D,x) 
\frac{\partial u}{\partial x} \Bigr) =0 , \ee 
where $v^2=F_0/\rho_0$ can be considered as the velocity. 
This equation is an equation for fractal solid string.

\section{Solution of Wave Equation for Fractal Solid String}

The wave equation for the fractal solid string has the following form
\be \label{sp} s(x) \frac{\partial^2 u}{\partial t^2}=
\frac{\partial}{\partial x}
\Bigl( p(x) \frac{\partial u}{\partial x}\Bigr) , \ee
where the functions $s(x)\ge 0$ and $p(x)\ge 0$ are defined by
\[ s(x)=c_1(D,x) , \quad p(x)=v^2 c_1(D,x) . \]
Let us consider the region $0\le x \le l$ and the following conditions
\[ u(x,0)=f(x), \quad \left(\frac{\partial u}{\partial t} \right)(x,0)=g(x), \]
\[ u(0,t)=0, \quad u(l,t)=0 . \]
The solution of Eq. (\ref{sp}) has the form
\[ u(x,t)=\sum^{\infty}_{n=1} \Bigl( f_n \cos(\lambda_n t)+
\frac{g_n}{\sqrt{\lambda_n}} \sin(\lambda_n t) \Bigr) y_n(x) . \]
Here $f_n$ and $g_n$ are the Fourier coefficients for the functions
$f(x)$ and $g(x)$:
\[ f_n=||y_n||^{-2} \int^l_0 f(x) y_n(x) d l_D =
||y_n||^{-2} \int^l_0 c_1(D,x) f(x) y_n(x) d x, \]
\[ g_n=||y_n||^{-2} \int^l_0 g(x) y_n(x) d l_D=
||y_n||^{-2} \int^l_0 c_1(D,x) g(x) y_n(x) dx, \]
where we use
\[ ||y_n||^2=\int^l_0 y^2_n(x) d l_D =\int^l_0 c_1(x,D) y^2_n(x) dx . \]
Note that the eigenfunctions $y_n(x)$ satisfy the following condition
\[ \int^l_0 y_n(x) y_m(x) d l_D=\delta_{nm} . \]

The eigenvalues $\lambda_n$ and the eigenfunctions $y_n(x)$
are defined as solutions of the equation
\[ v^2[c_1(D,x)y^{\prime}_x]^{\prime}_x+\lambda^2 c_1(D,x) y=0, \quad
y(0)=0, \quad y(l)=0 , \] 
where $y^{\prime}_x=dy(x)/dx$. 
This equation can be rewritten in an equivalent form
\be \label{eq} v^2 xy^{\prime \prime}_{xx}(x)+(D-1)y^{\prime}_{x}(x)
+ \lambda^2 x y(x)=0 . \ee
The solution of Eq. (\ref{eq}) has the form
\[ y(x)=C_1x^{1-D/2}J_{\nu}(\lambda x/v)+
C_2 x^{1-D/2} Y_{\nu}(\lambda x/v) , \] 
where $\nu= |1-D/2|$. 
Here $J_{\nu}(x)$ are the Bessel functions of the first kind,
and $Y_{\nu}(x)$ are the Bessel functions of the second kind.

As an example, we consider the case that is defined by
\[ l=1, \quad v=1, \quad 0\le x \le 1, \quad f(x)=x(1-x), \quad g(x)=0 . \]

The usual string has $D=1$ and the solution is defined by 
\[ u(x,t)=\sum^{\infty}_{n=1} \frac{4(1- (-1)^n) 
\sin(\pi n x) \cos (\pi n t) }{\pi^3 n^3} . \]
The approximate solution for the usual string with $D=1$ that has the form
\[ u(x,t) \simeq \sum^{10}_{n=1} \frac{4(1- (-1)^n) 
\sin(\pi n x) \cos (\pi n t) }{\pi^3 n^3}  \]
is shown in Fig. 1 for $0\le t \le 3$ and the velocity $v=1$.

\begin{figure}[tbh]
\begin{center}
\resizebox{11cm}{!}{\includegraphics{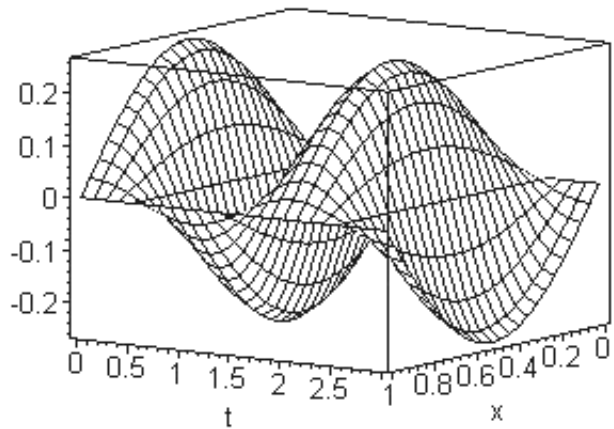}}
\caption{\it Usual wave (D=1) with the velocity $v=1$.}  
\label{US10}
\end{center}
\end{figure}

The fractal solid string with $D=1/2$, has the functions
$y_n(x)$ in the form 
\[ y_n(x)=\frac{1}{\Gamma(1/2)} x^{3/4} J_{3/4} (\sqrt{2} \lambda_n x/2) . \]
The eigenvalues $\lambda_n$ are the zeros of the Bessel function
\[ \lambda_n: \quad J_{3/4} (\sqrt{2} \lambda_n /2)=0 . \] 
For example,  
\[ \lambda_1 \simeq 4.937, \quad \lambda_2 \simeq 9.482, \quad
 \lambda_3 \simeq 13.862, \quad \lambda_4 \simeq 18.310, \quad
 \lambda_5 \simeq 22.756 . \]
The approximate values of the eigenfunctions
\[ f_n=\frac{||y_n||^{-2}}{\Gamma(D)} \int^l_0 x^{5/4} (1-x) 
J_{3/4} (\sqrt{2} \lambda_n /2) dx  , \]
are following
\[ f_1 \simeq 1.376, \quad f_2 \simeq -0.451, \quad f_3 \simeq 0.416, \quad
f_4 \simeq -0.248, \quad f_5 \simeq 0.243. \]

The solution of the wave equation for the fractal solid string 
with $D=1/2$ is
\[ u(x,t)=\sum^{\infty}_{n=1} f_n \cos(\lambda_n t) 
J_{3/4}(\sqrt{2} \lambda_n x /2) .\]
The approximate solution for the fractal solid string with 
the mass dimension $D=1/2$ has the form
\[ u(x,t) \simeq 
\sum^{10}_{n=1} f_n \cos(\lambda_n t) J_{3/4}(\sqrt{2} \lambda_n x /2)  \]
is shown in Fig. 2 for the velocity $v=1$.

\begin{figure}[tbh]
\begin{center}
\resizebox{11cm}{!}{\includegraphics{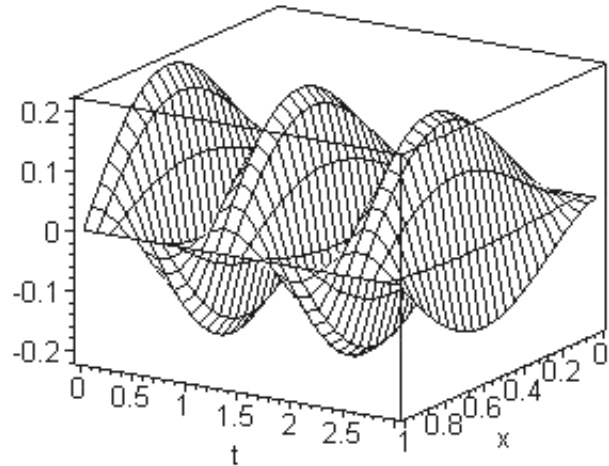}}
\caption{\it Wave on fractal solid string with $D=1/2$ 
and the velocity $v=1$.}  
\label{FS10}
\end{center}
\end{figure}

\section{Conclusion}

The fractional continuous models for fractal solid 
can have a wide application. 
This is due in part to the relatively small numbers of parameters 
that define a fractal solid of great complexity and rich structure.
In many cases, the real fractal structure of matter can be disregarded 
and the fractal solid can be replaced by some fractional 
continuous model \cite{PLA05,AP05-2}. 
In order to describe the solid with 
non-integer mass dimension, we must use the fractional calculus.
Smoothing of the microscopic characteristics over the 
physically infinitesimal volume transforms the initial 
fractal solid into fractional continuous model
that uses the fractional integrals. 
The order of fractional integral is equal 
to the fractal mass dimension of the solid.
The fractional continuous model allows us to describe dynamics of 
wide class fractal media \cite{AP05-2,PLA05-2,Chaos05,Physica,JPA05-2}.



\end{document}